\documentclass[a4paper]{JAC2003}
\usepackage{graphicx}
\usepackage{booktabs}
\usepackage{subfig}
\usepackage{amsmath}
\usepackage{url}

\usepackage{varwidth}
\usepackage{xcolor}

\newcommand{\betastar}{\ensuremath{\beta^{*}}}

\begin{document}
\title{OPERATIONAL CONSIDERATIONS ON THE STABILITY OF COLLIDING BEAMS}

\author{X. Buffat, EPFL, Lausanne, Switzerland; CERN, Geneva, Switzerland \\
W. Herr, T. Pieloni, CERN, Geneva, Switzerland}

\maketitle

\begin{abstract}
While well studied in the absence of beam--beam and while colliding head-on, the stability of the LHC beams can be very critical in intermediate steps. During the squeeze, the long-range beam--beam interaction becomes a critical component of the beam's dynamics. Also, while the transverse separation at the interaction points is collapsed, the beam--beam forces change drastically, possibly deteriorating the beam's stability. Finally, during luminosity production, the configuration of the LHC in 2012 included few bunches without head-on collision in any of the interaction points having different stability properties. Stability diagrams are being evaluated numerically in these configurations in an attempt to explain instabilities observed in these phases during the 2012 proton run of the LHC.
\end{abstract}

\section{Introduction}
The LHC configuration changes significantly along a standard operational cycle. These different configurations have different implications from the point of view of beam stability; in particular, the effect of Beam--Beam (BB) interactions can be very different. The approach described in~\cite{sdiag} is used to derive stability diagrams in the configurations encountered during the LHC run 2012 and the results are compared to the observations.

\section{Betatron squeeze}

\begin{figure}
 \centering
\includegraphics[width=0.7\linewidth]{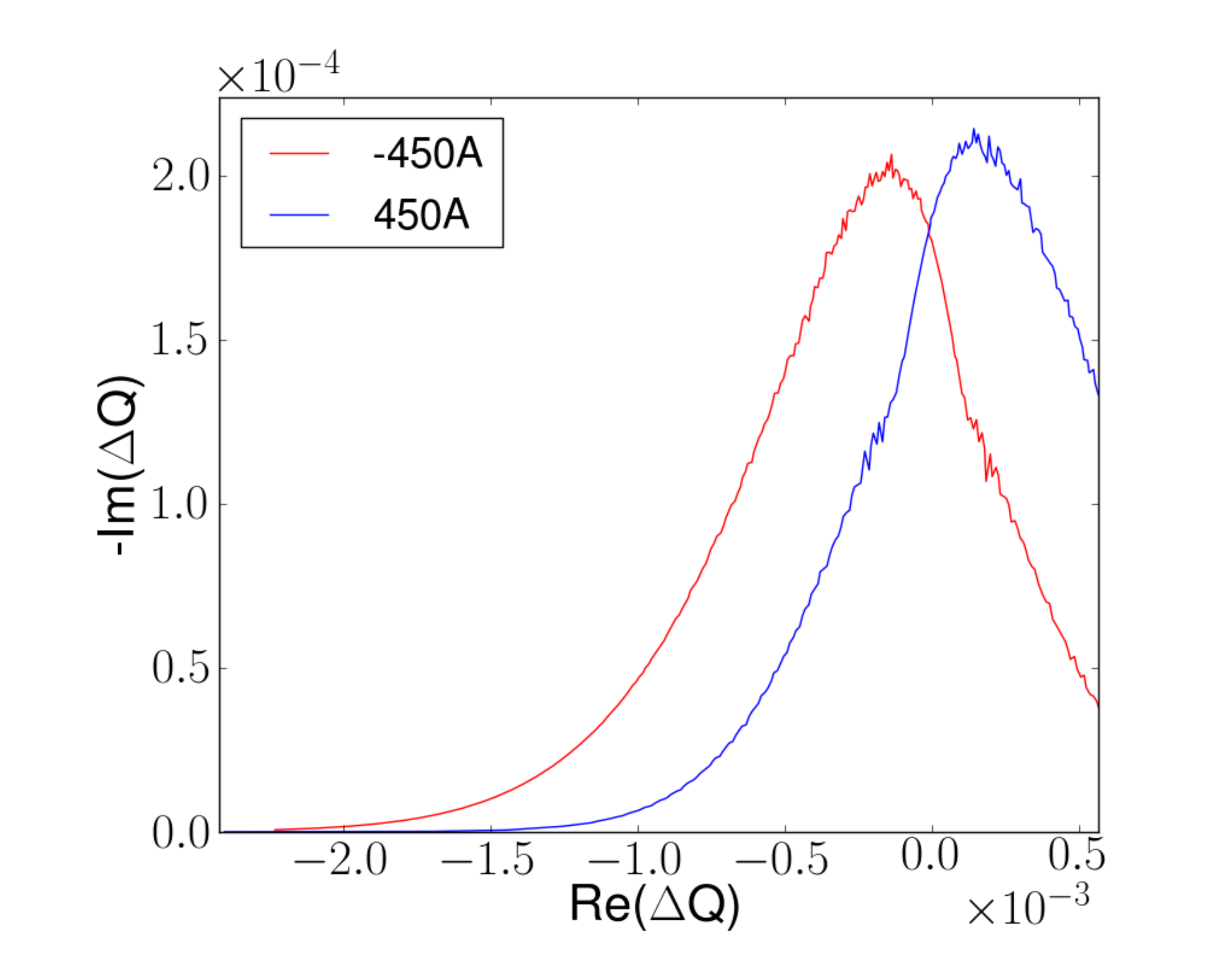}
\caption{Stability diagrams from octupoles with both polarities.}
\label{fig-octupoles}
\end{figure}

\begin{figure}
 \centering
\includegraphics[width=0.7\linewidth]{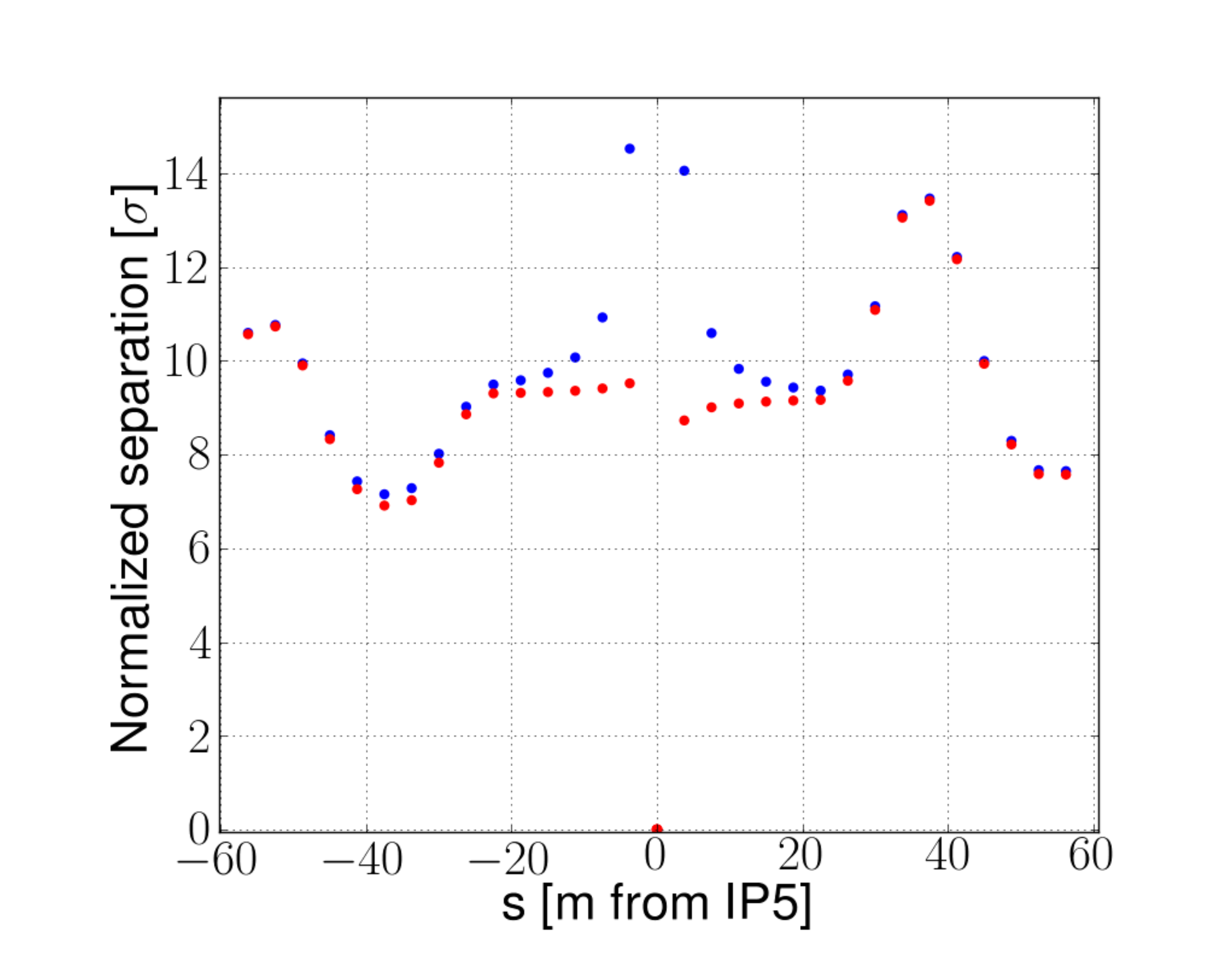}
\caption{Normalized beam--beam separation in IP5 at the end of the squeeze (blue) and in collision (red).}
\label{fig-sep eos}
\end{figure}

\begin{figure}
 \centering
\includegraphics[width=1.0\linewidth]{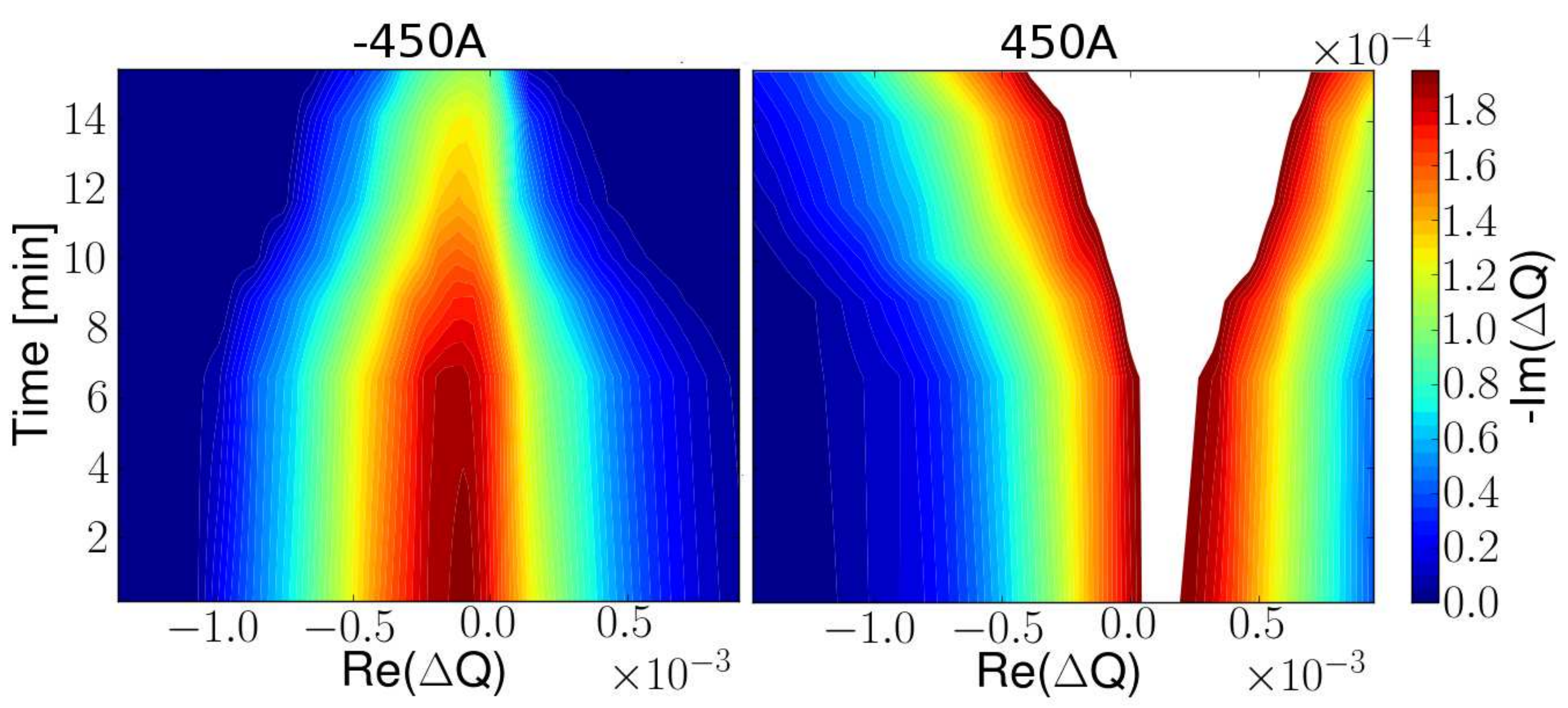}
\caption{Stability diagram as a function of time during the squeeze for both octupole polarity ($\pm450$~A). The \betastar s at $t=0$ are 11~m in IP1\&5 and 10~m in IP2\&8, at the end 0.6~m and 3~m respectively. This represents the most common bunch, with the largest number of LRBB interactions; the effect is similar but of lower amplitude for bunches with a lower number of LRBB.}
\label{fig-colorsdiag squeeze}
\end{figure}

\begin{figure}
 \centering
\subfloat[BBQ]{
\includegraphics[width=0.7\linewidth]{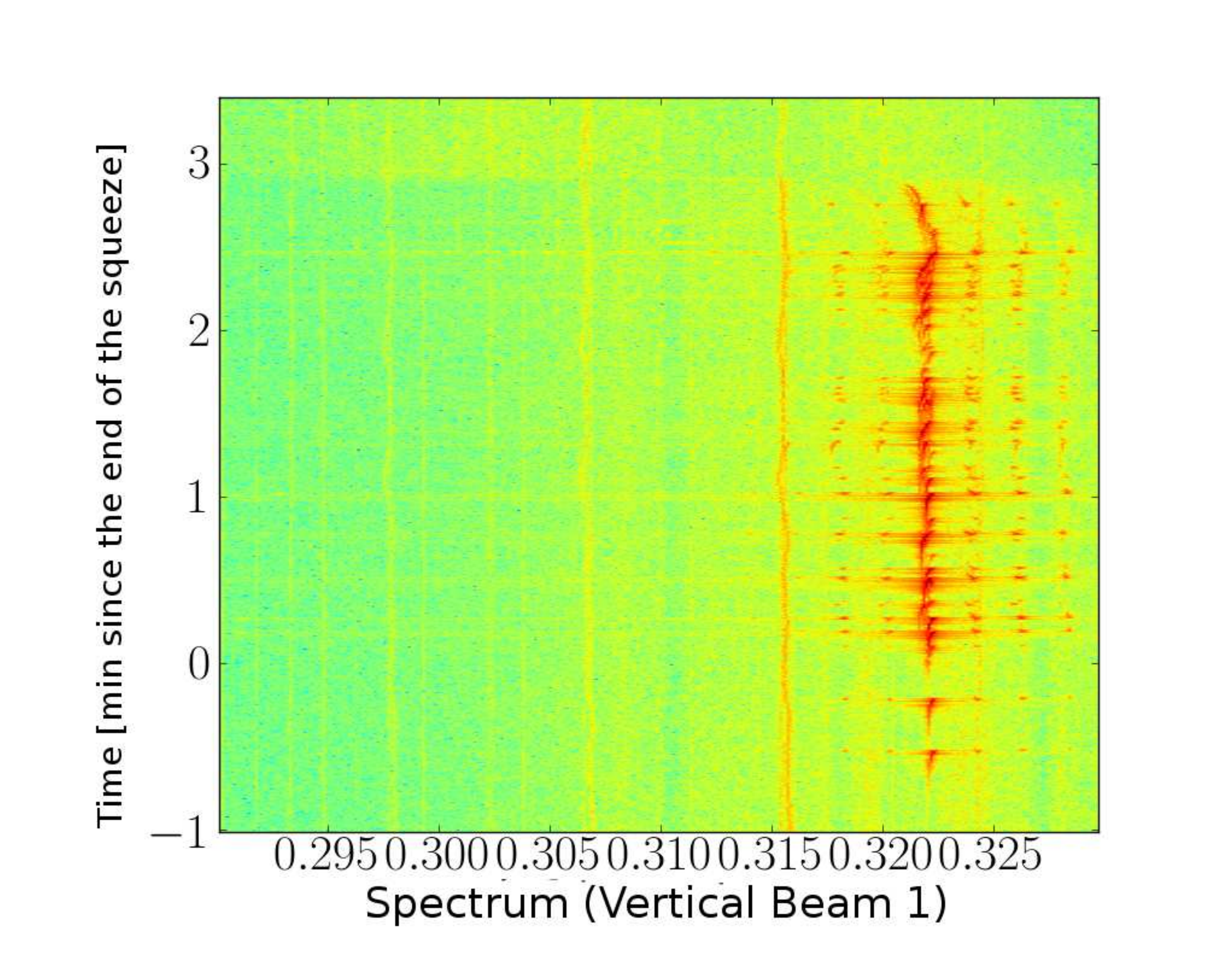}
}
\qquad
\subfloat[FBCT] {
\includegraphics[width=0.7\linewidth]{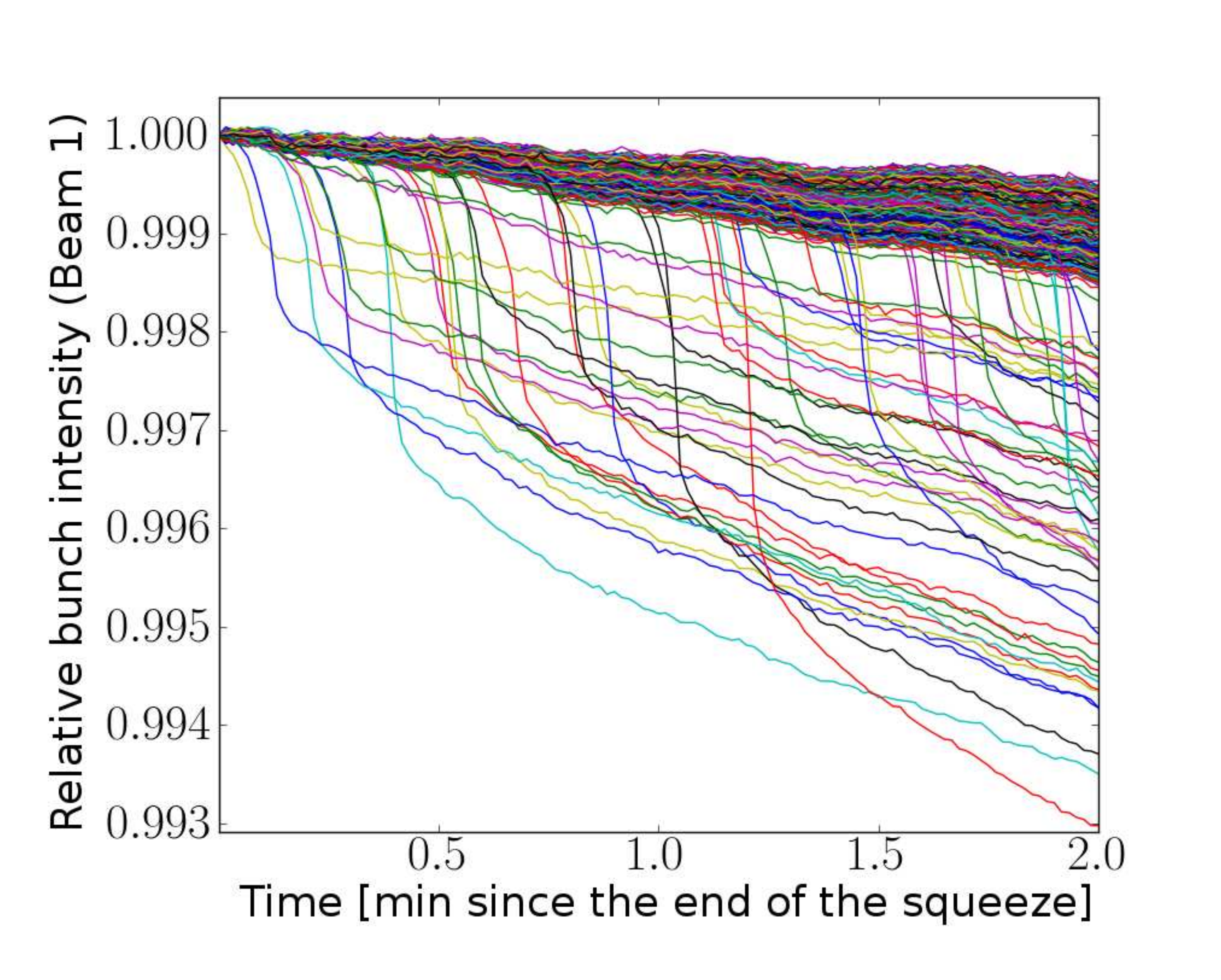}
}
\caption{Typical observation of an instability at the end of the squeeze during fill 3250. The machine is filled with 1374 bunches per beam with $\sim1.6\cdot10^{11}$ protons per bunch and emittances of $\sim 2.4\cdot10^{-6}~\mu$~m. The chromaticities are set to $\sim10$ units, the transverse feedback gain to 50 turns and the octupoles powered with $533$A. From $t=1.8$ to $2.9$ the beams are being brought to collision and are fully stabilized once colliding head-on.}
\label{fig-eos instability}
\end{figure}

Before the squeeze, BB interactions can be neglected. The stability is ensured by the transverse damper and amplitude detuning from the octupoles. They can be powered with up to $\sim500$~A, with either polarity. The resulting stability diagrams for each polarity are shown in Fig.~\ref{fig-octupoles}. As the expected tune shifts in the LHC have negative real parts~\cite{tuneshifts}, the negative polarity is preferable in this configuration and therefore was chosen as the design value. However, going through the squeeze, the effect of the Long-Range BB (LRBB) encounters starts playing a significant role. As can be seen in Fig.~\ref{fig-sep eos}, at the end of the squeeze, most of the LRBB interactions are already at the separation at which they will stay during luminosity production, the only difference being the separation orbit bump. As can be seen in Fig.~\ref{fig-colorsdiag squeeze}, the stability diagram changes dramatically during the squeeze, in particular, the negative polarity is no longer preferable. Some instabilities at the end of the squeeze were attributed to this compensation and consequently the polarity was changed~\cite{polarity change}. The benefit from the change of polarity could not be properly assessed as this change in the operational configurations appear alongside a large increase of the chromaticity, from 2 to 15 units, and the transverse feedback gain, from more than 100 turns to 50 turns. While these stabilizing techniques have allowed the machine performance to be increased, by reducing the number of dumps due to beam losses caused by coherent instabilities, they have not cured the instability as it was still clearly visible (Fig.~\ref{fig-eos instability}). In this new configuration, however, it is clear that the modification of the tune spread due to LRBB can not explain the instabilities observed, as the stability diagram is larger at the end, with respect to the beginning of the squeeze at which the beams are stable.



Several investigations are currently ongoing to understand the instabilities at the end of the squeeze. In particular, the stability diagrams presented are not suited to describing the stability of multibunch modes in the presence bunch dependent amplitude detuning, nor are they suited to coherent beam--beam modes. These effects are currently studied using multiparticle tracking~\cite{simon}. Other effects are also under study, such as external noise~\cite{sdiag} or optics imperfections.

As in Fig.~\ref{fig-eos instability}, it has been observed that the instability at the end of the squeeze is always well stabilized once the beams are colliding head-on, therefore it is considered to go through the squeeze with colliding beams in future scenarios~\cite{colliding squeeze}. As discussed in the section on luminosity production, this approach not only offers a cure for the instability, but also provides a significant margin for increased impedance or beam brightness.

\section{Bringing the beams into collision}
\begin{figure}
 \centering
\includegraphics[width=0.7\linewidth]{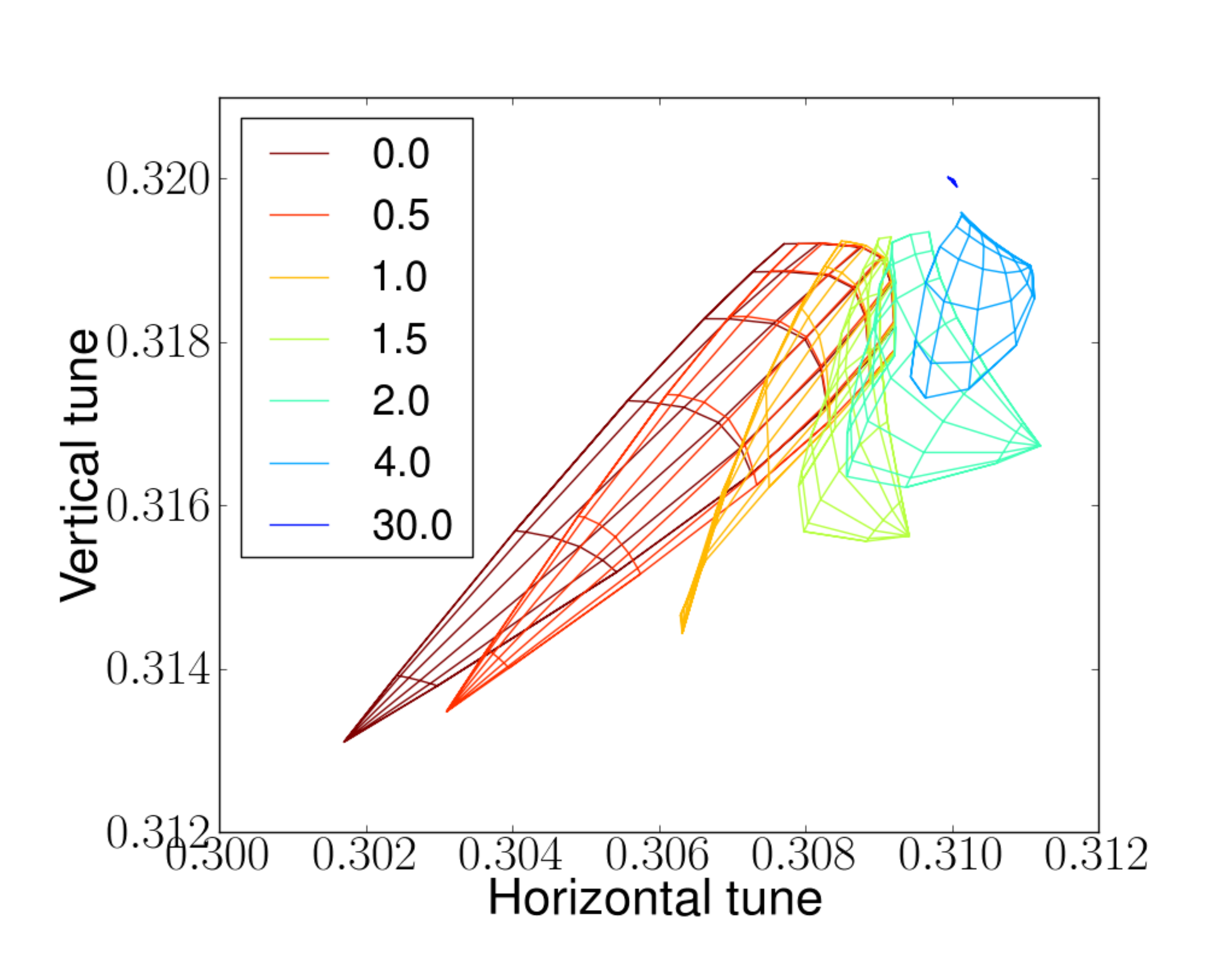}
\caption{Example of tune footprint of a bunch colliding in IP1 with different separations in the horizontal plane.}
\label{fig-footprints sepscan}
\end{figure}

\begin{figure}
 \centering
\subfloat[With LRBB]{
\includegraphics[width=1.0\linewidth]{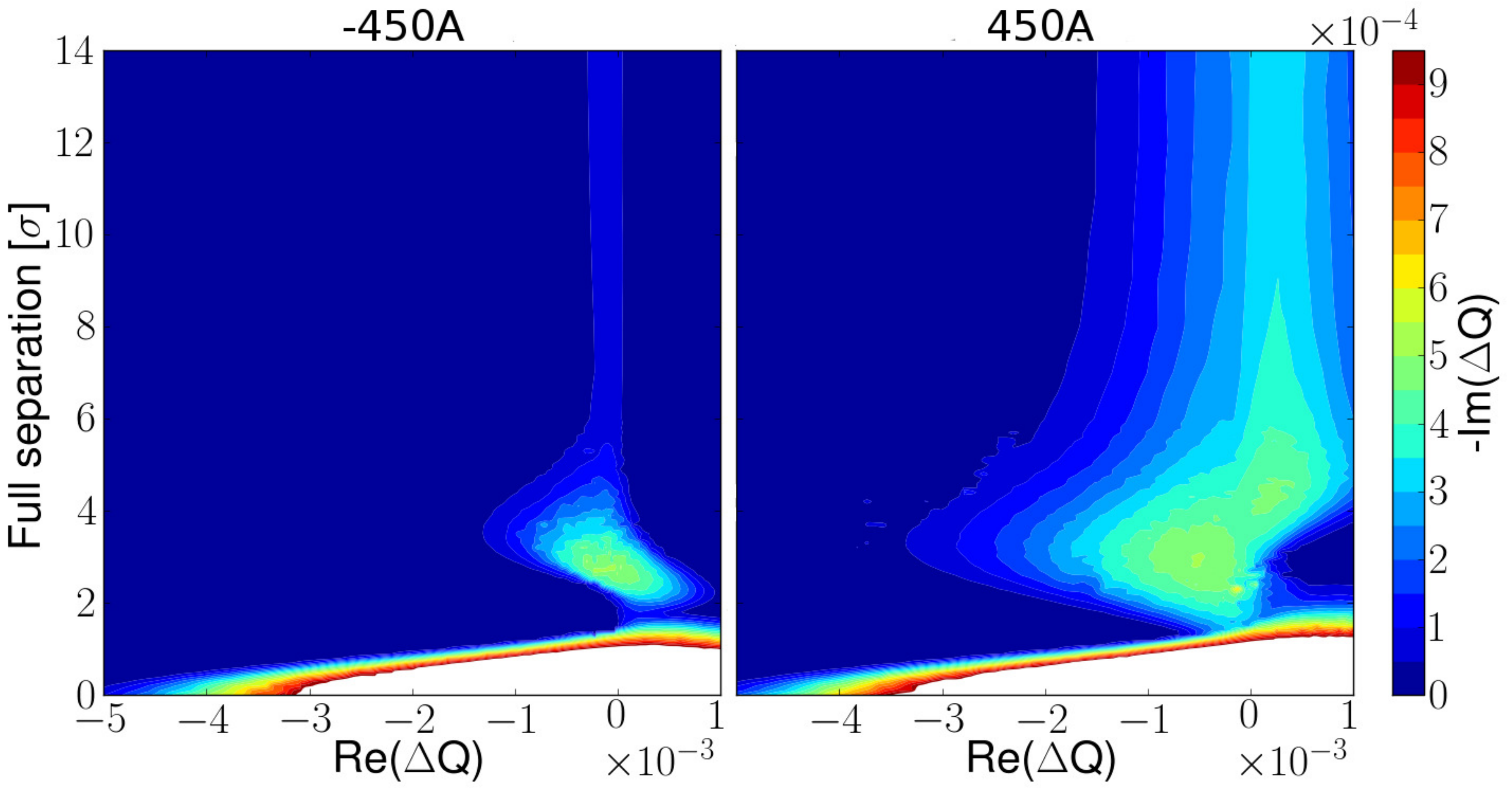}
\label{fig-sdiag sepscan}
}
\qquad
\subfloat[Without LRBB]{
\includegraphics[width=1.0\linewidth]{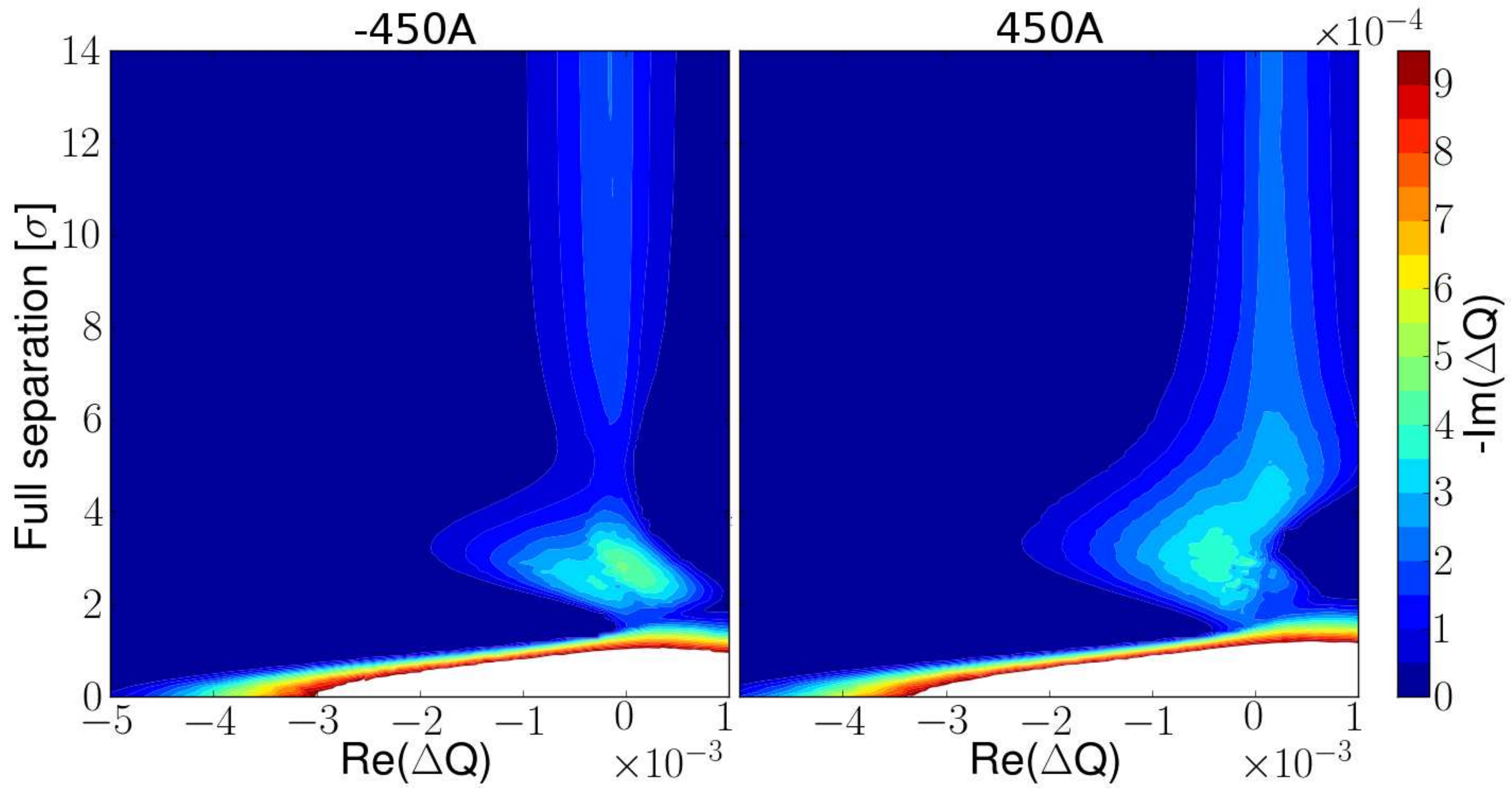}
\label{fig-sdiag sepscan nolr}
}
\caption{Stability diagram as a function of beam separation in IP1\&5 for a bunch with either maximum number of LRBB or none, and both polarities of the octupoles.}
\label{fig-sdiag}
\end{figure}

When the parallel separation is collapsed, in order to bring the beams into collision the tune shift and spread of the colliding bunches change sign as illustrated by Fig.~\ref{fig-footprints sepscan}, leading to a significant modification of the stability diagram. As shown by Fig.~\ref{fig-sdiag}\subref{fig-sdiag sepscan}, the stability diagram is enhanced for separation in the order of 2 to 4$\sigma$ and drastically reduced around 1.5$\sigma$. This minimum of stability depends significantly on the configuration considered and therefore can be very different for bunches having different numbers of LRBB or Head-On BB (HOBB). In this case, the reduction of the stability diagram is however not due to a compensation of tune spread as at the end of the squeeze, it is caused by a change of sign of the tune spread which leads to a systematic cancellation of nearby poles in the dispersion integral. Even if the minimum stability also exists, it is clear from Fig.~\ref{fig-sdiag}\subref{fig-sdiag sepscan} that the positive polarity of the octupole is also favourable in this configuration. One should however not forget that the stability must be ensured for all bunches; in particular, in most LHC configurations there exist some bunches with very few LRBB; the minimum stability for these bunches can still be very critical, as shown by Fig.~\ref{fig-sdiag}\subref{fig-sdiag sepscan nolr}.


\begin{figure}
 \centering
\includegraphics[width=0.7\linewidth]{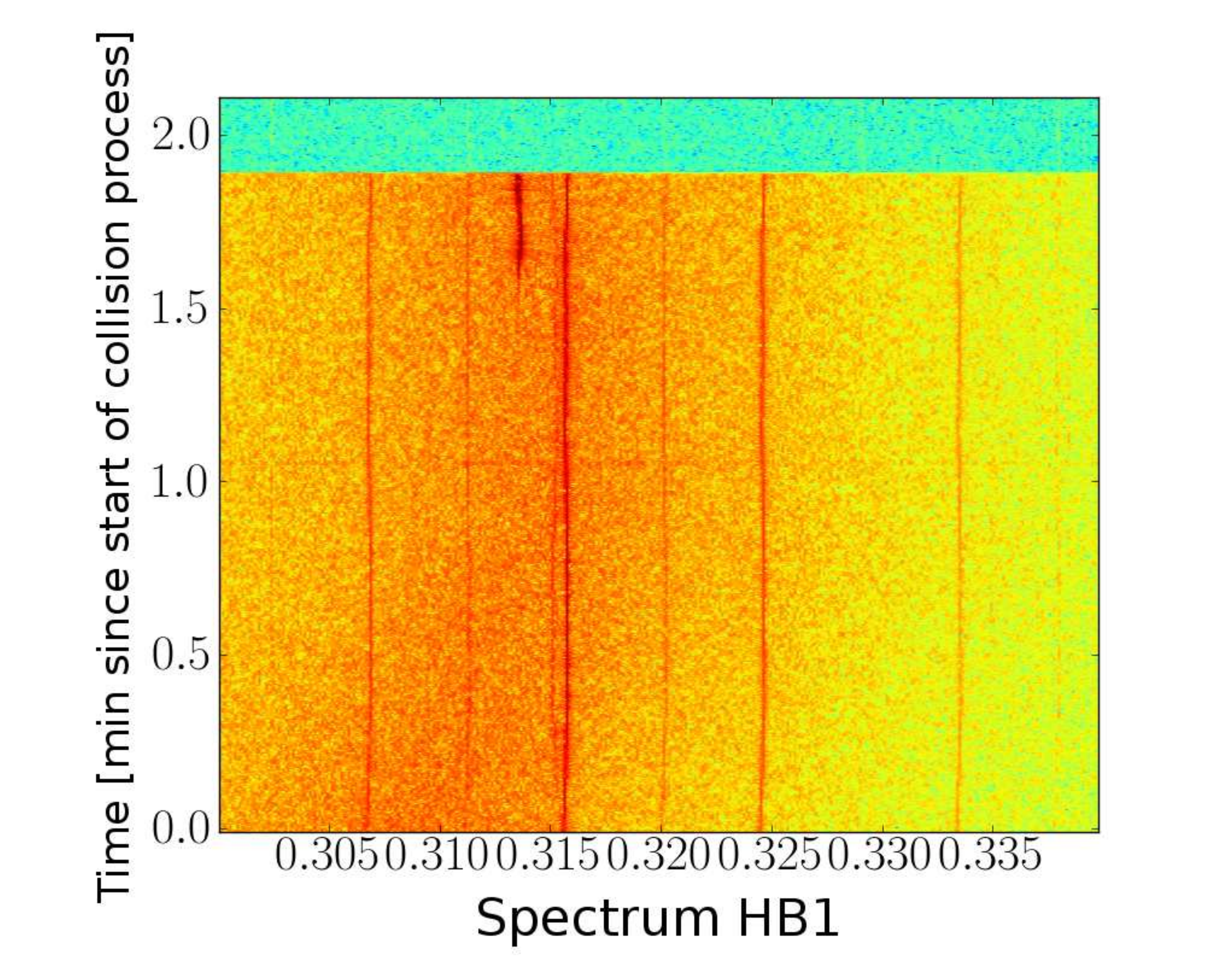}
\caption{Spectrogram measured by the BBQ in the horizontal plane of Beam 1 during the collision process of fill 2808 (i.e. old implementation). An instability is visible at a time corresponding to separations around $2.3\sigma$, the beams are then dumped due to beam losses.}
\label{fig-BBQ spectrogram 2808}
\end{figure}

\begin{figure}
 \centering
\subfloat[Old process] {
\includegraphics[width=0.7\linewidth]{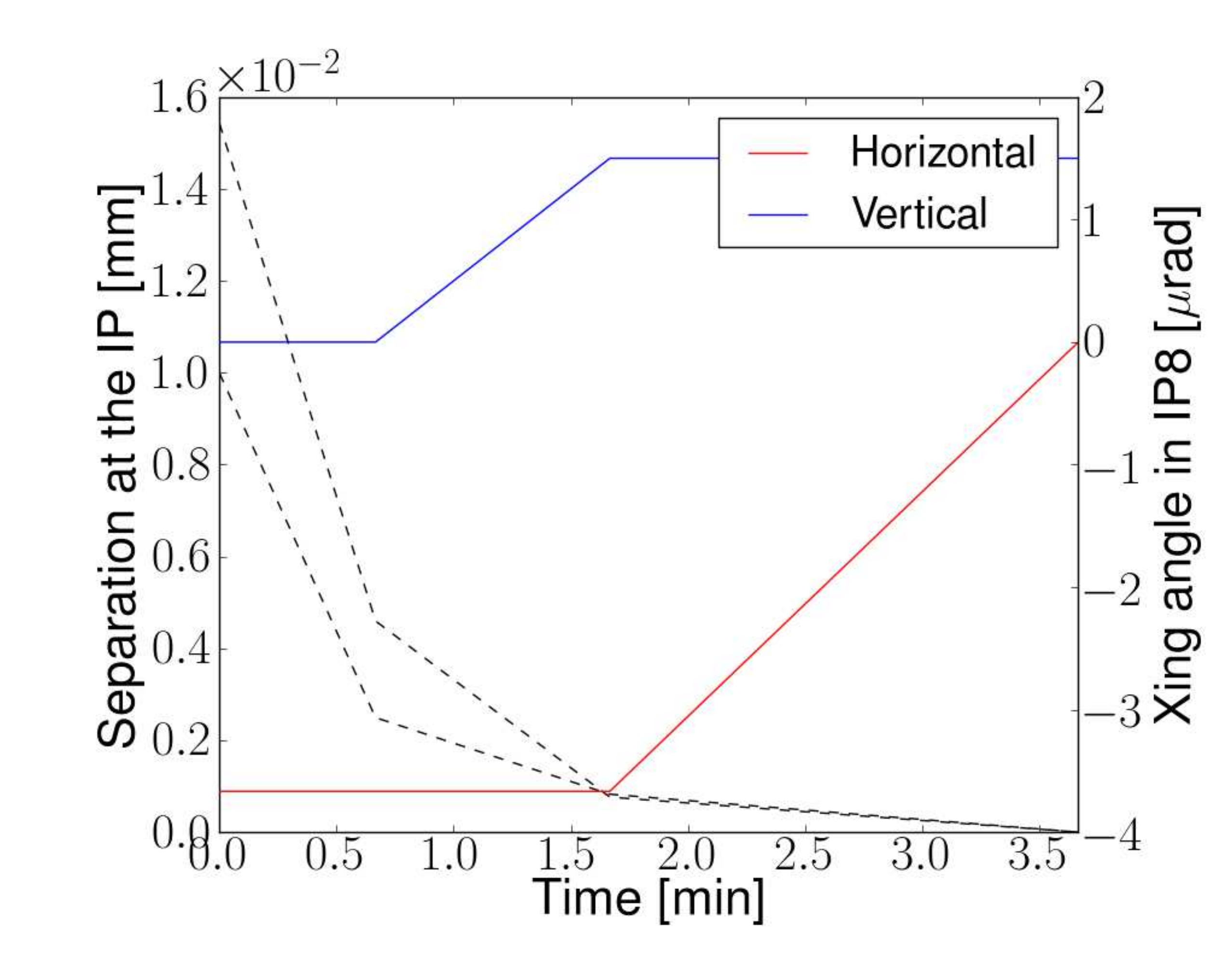}
\label{fig-adjust old}
}
\qquad
\subfloat[New process] {
\includegraphics[width=0.7\linewidth]{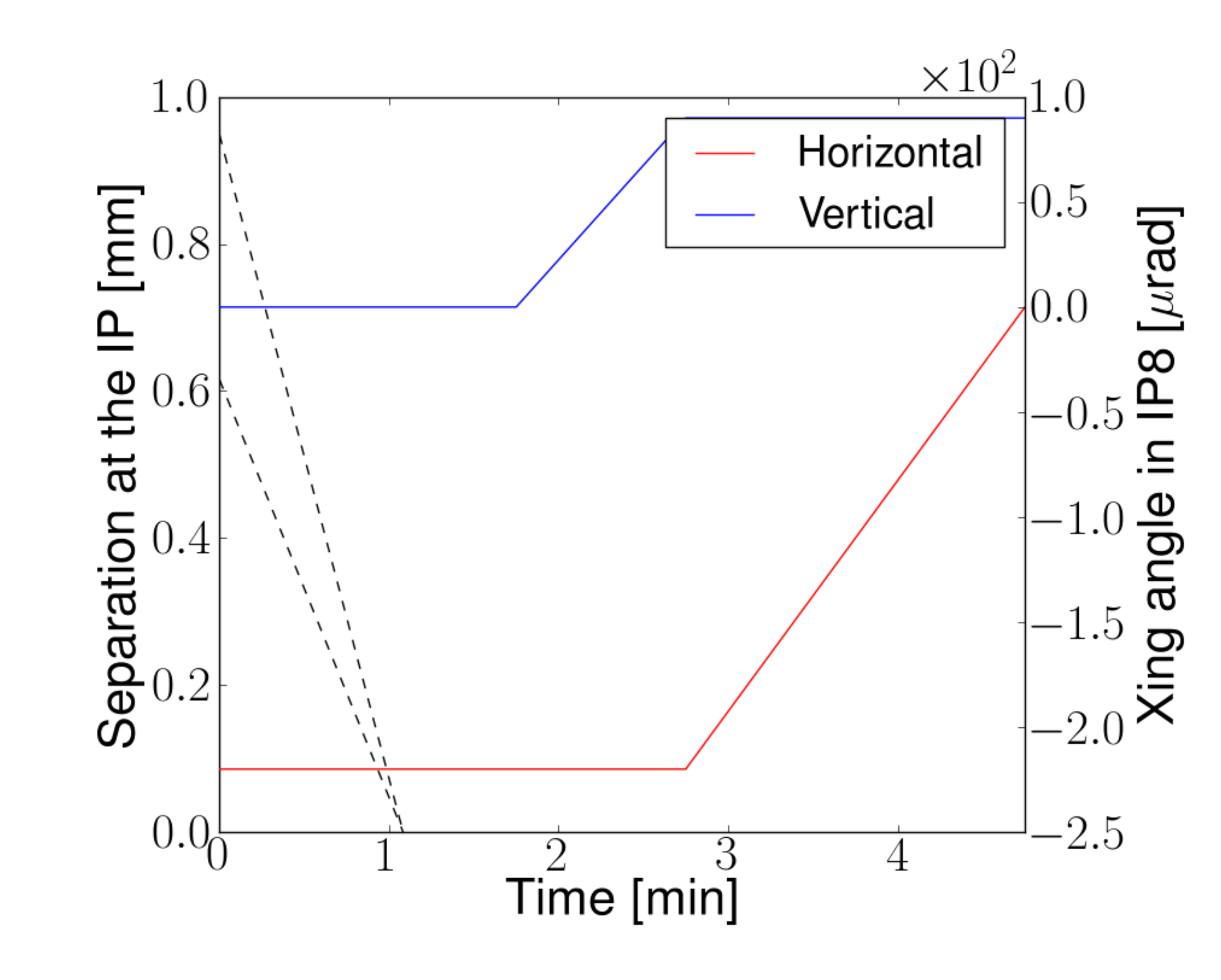}
\label{fig-adjust new}
}
\caption{Two implementations of the process that brings the beams into collision and tilts the Xing angle in IP8.}
\label{fig-adjust}
\end{figure}

There have been several observations of coherent instability during the process that brings the beams into collision during the 2012 run of the LHC, a spectrogram during such instability is shown in Fig.~\ref{fig-BBQ spectrogram 2808}. The separations at which these instabilities occur is in qualitative accordance with the critical separation discussed above. It is however difficult to make quantitative comparison as many critical observables are not available with sufficient accuracy, such as chromaticities and bunch by bunch emittances. While small separations may be very critical in term of stability, it did not prevent collision in previous years. In addition to the increased impedance due to tighter collimator settings and increased beam brightness, a critical change is the implementation of the process that brings the beams into collision. As can be seen in Fig.~\ref{fig-adjust}\subref{fig-adjust old}, the implementation of this process included, in early 2012, a change of the crossing angle in IP8~\cite{IP8 tilting}, resulting in an extended time spent at critical separations. This could be avoided by a change in the implementation of the process that brings the beams to head-on collision as fast as possible before going through other manipulation (Fig.~\ref{fig-adjust}\subref{fig-adjust new}). Other cures to such instability exist; multiparticle tracking simulations suggest that they are well damped by high positive chromaticity or high transverse feedback gain (thanks to a private communication by S. White in 2012). In particular, such instabilities were no longer observed in the LHC after a change of configuration to high chromaticity, high damper gain and positive polarity of the octupole. The short process could only be tested in this new configuration; there would be, however, an interest from the beam lifetime point of view in being able to run with lower chromaticity and damper gain, which, in this case, may be achieved by speeding up the collision process.

\begin{figure}
 \centering
\includegraphics[width=0.9\linewidth]{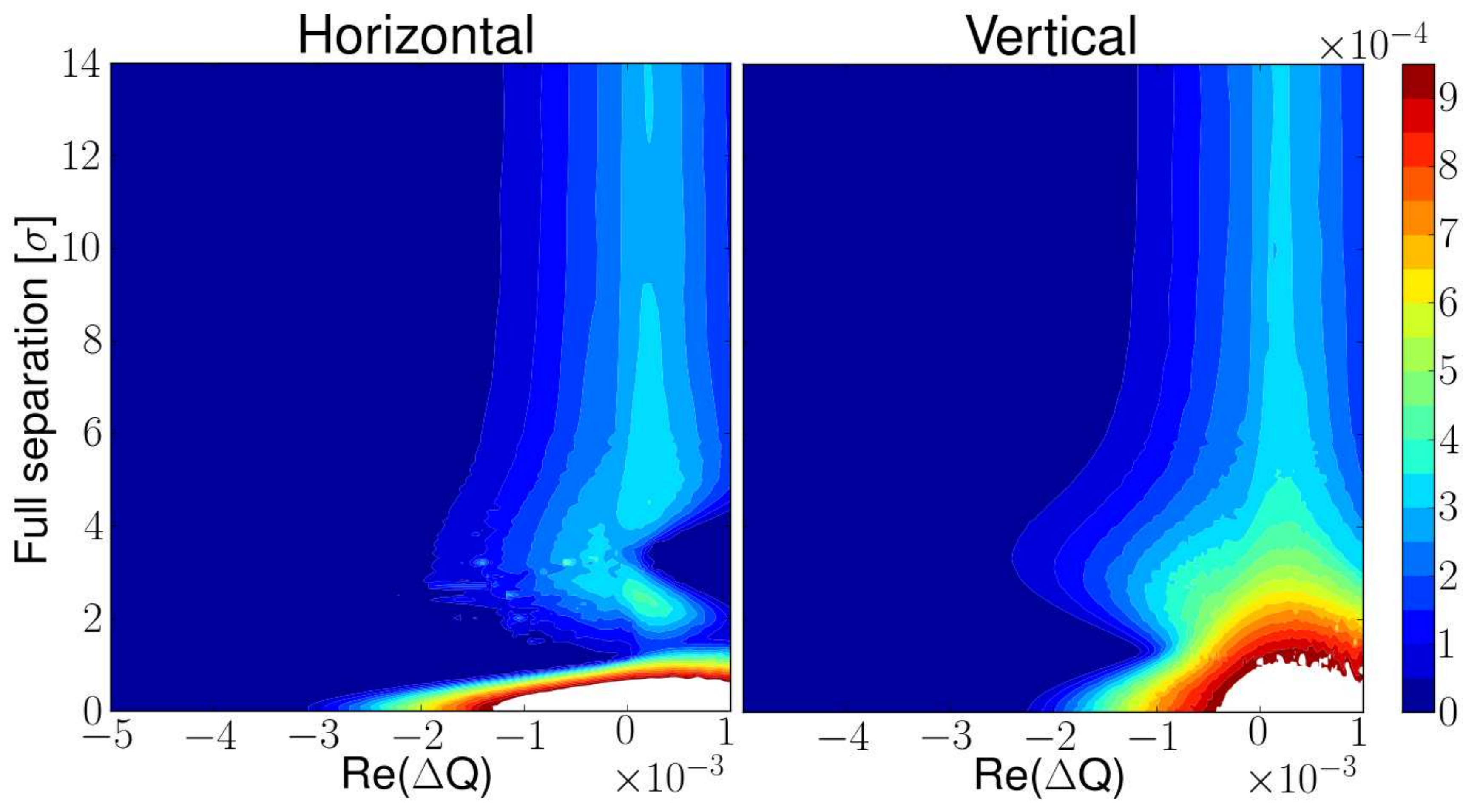}
\caption{Stability diagram while collapsing the separation in IP1 only (horizontal separation).}
\label{fig-sdiag adjust IP1}
\end{figure}

In case this should not suffice, the possibility to go into collision one after the other may be interesting. Indeed, as can be observed in Fig.~\ref{fig-sdiag adjust IP1}, in this configuration the minimum stability is reached in one plane only. Whereas coupling is assumed to be negligible in our approach, simulation studies suggest that the stability of the two planes could be shared via non-linear coupling of the beam--beam force. It is important to note that even though the beams are separated in one plane only in the model, the machine imperfections create a separation in the other plane. In this configuration, it is important to keep this separation well corrected, as a separation in both planes at one IP would result in a situation similar to both IP1\&5 simultaneously.

\section{Luminosity production} \label{lumiprod}
\begin{figure}
 \centering
\includegraphics[width=0.7\linewidth]{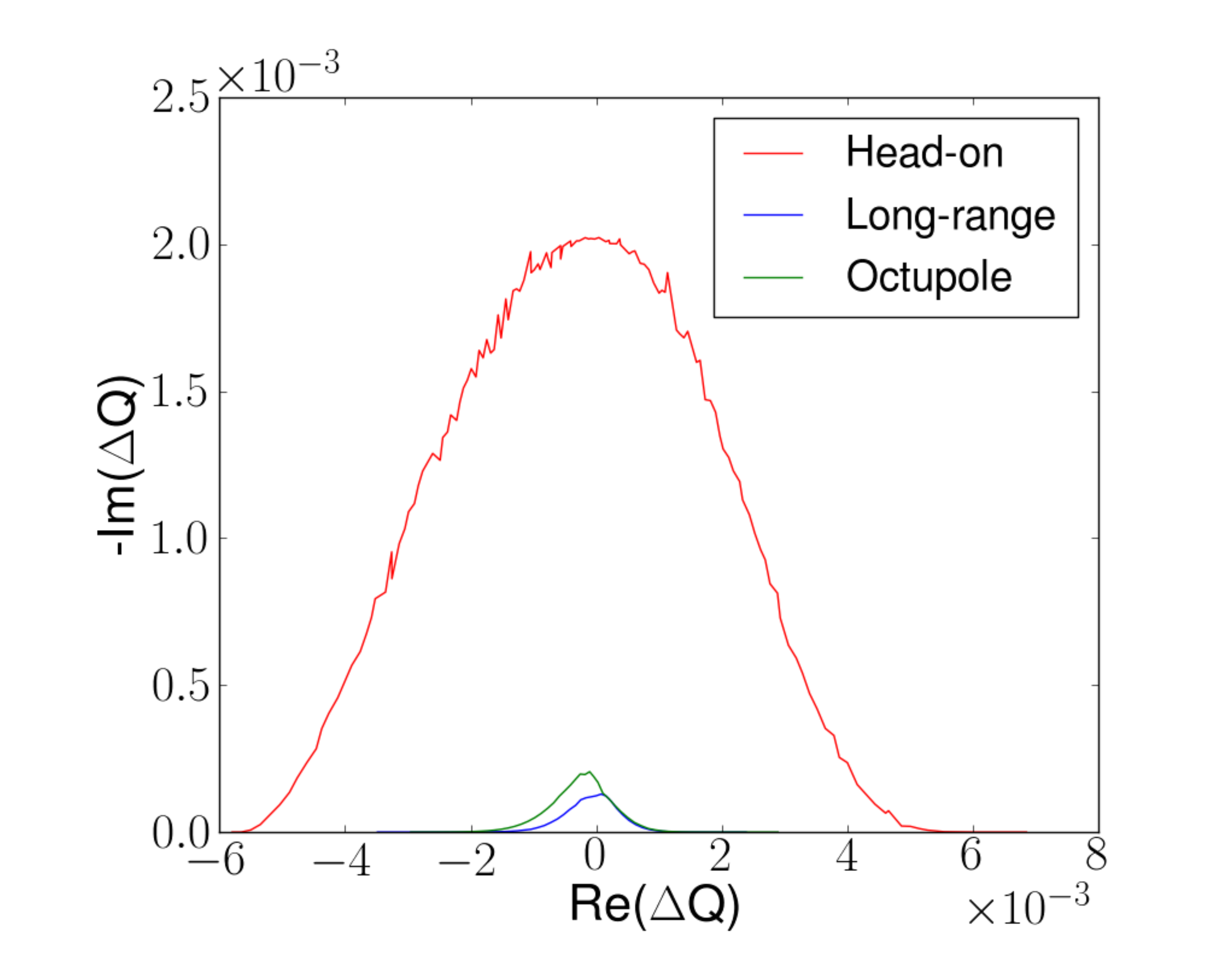}
\caption{Comparison of stability diagrams from either octupoles powered with -450~A, LRBB in IP1\&5 or HOBB in IP1\&5.}
\label{fig-sdiag HO}
\end{figure}
\begin{figure}
 \centering
\includegraphics[width=0.7\linewidth]{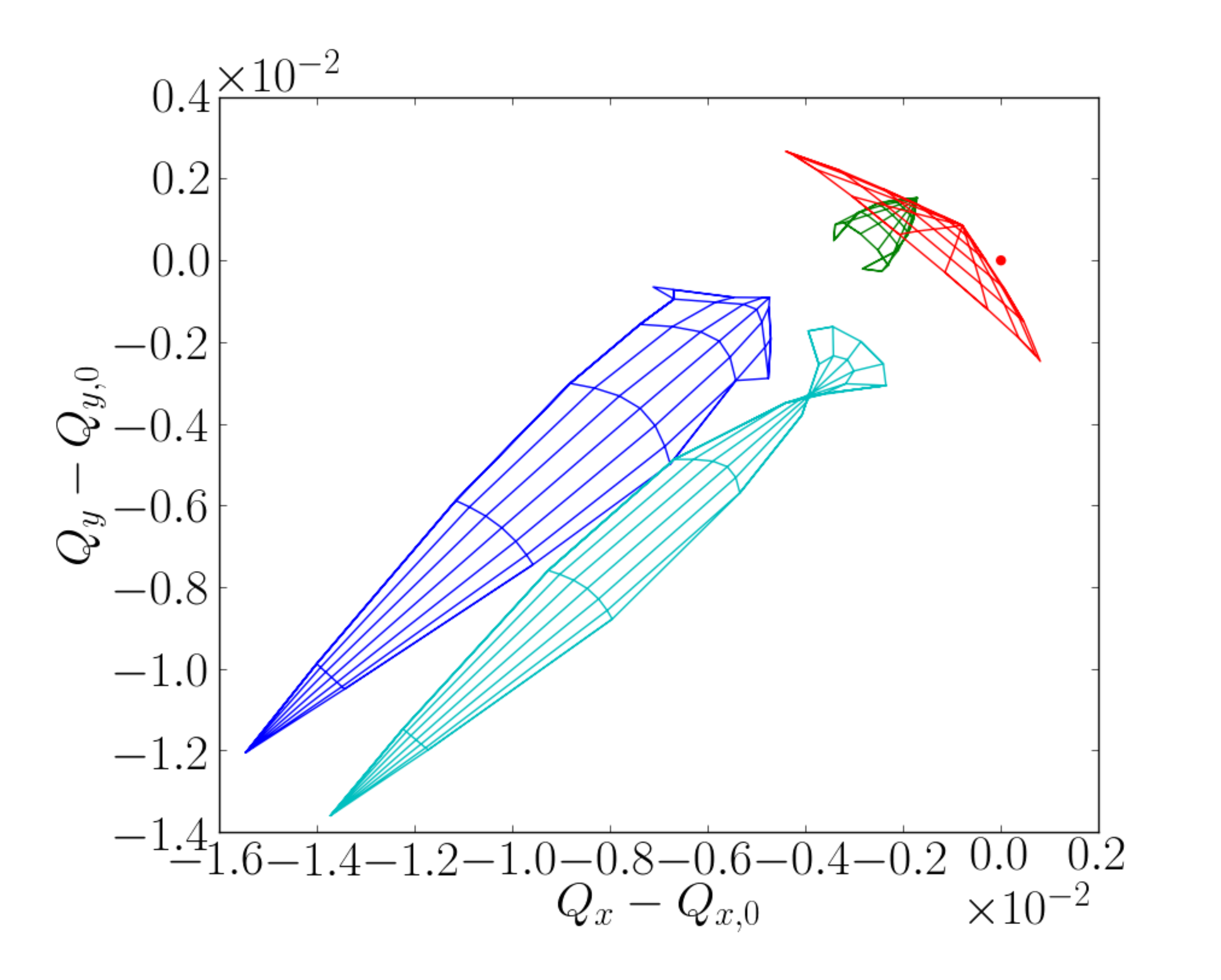}
\caption{Example of tune footprints of different bunches present simultaneously in the machine during luminosity production.}
\label{fig-stable beam foots}
\end{figure}
While colliding head-on, beam--beam is nominating the non-linearities undergone by the core of the beam and consequently provides the dominant contribution to the stability diagram. Fig.~\ref{fig-sdiag HO} compares stability diagrams from octupole, long-range and head-on; it is clear that HOBB collision is extremely efficient to providing stability, to the point that the stabilization techniques required before bringing the beams into collision are no longer required during luminosity production. This was however not so simple in the LHC configuration used in 2012. Indeed, luminosity in IP2 was provided by bunch-satellite collisions, which lead to an essentially inexistent HOBB contribution and IP8 luminosity was being levelled with a transverse offset. Therefore the only full HOBB collisions were in IP1\&5, where non-colliding bunches are requested~\cite{richard}. The complexity of this configuration is illustrated by Fig.~\ref{fig-stable beam foots}, representing the tune footprints of different bunches existing simultaneously in the machine during luminosity production. The stability of each bunch is crucial as the loss of part of a single bunch is enough to create a dump of the whole beam. This enforces the usage of strong stabilizing techniques, in particular high chromaticity, high transverse feedback gain and high octupole current, during luminosity production in order to stabilize bunches without head-on collision. In order to further optimize luminosity lifetime, it would be advisable to run in a configuration with one head-on collision for each bunch, allowing relaxation of the use of stabilizing techniques which are potentially harmful for the intensity lifetime and emittance growth of all bunches.

\subsection{Levelling with a Transverse Offset}
\begin{figure}
 \centering
\subfloat[Bunch intensity of IP8 private bunches]{
\includegraphics[width=0.7\linewidth]{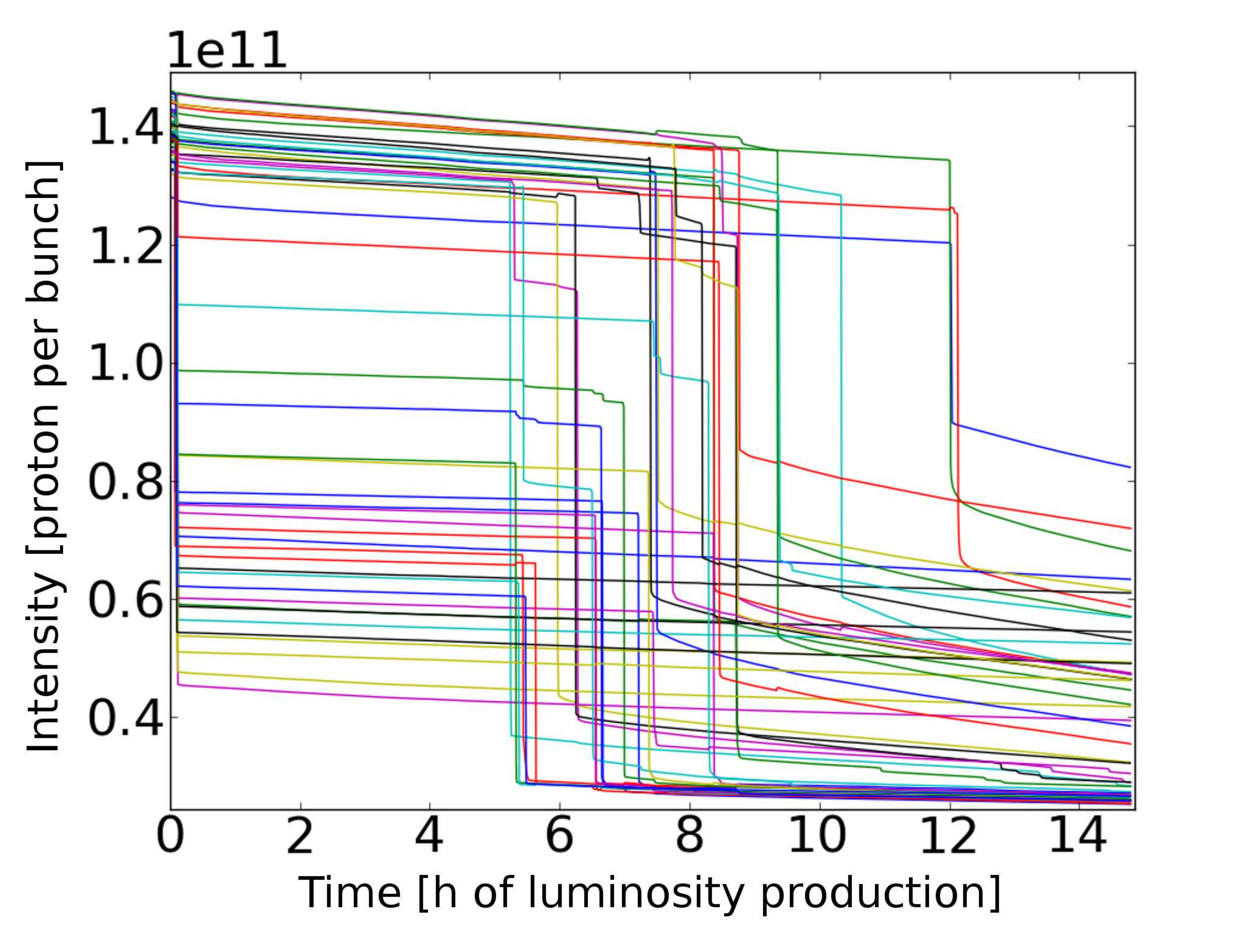}
\label{fig-snowgflake BCT}
}
\qquad
\subfloat[Relative bunch losses during luminosity production]{
\includegraphics[width=0.7\linewidth]{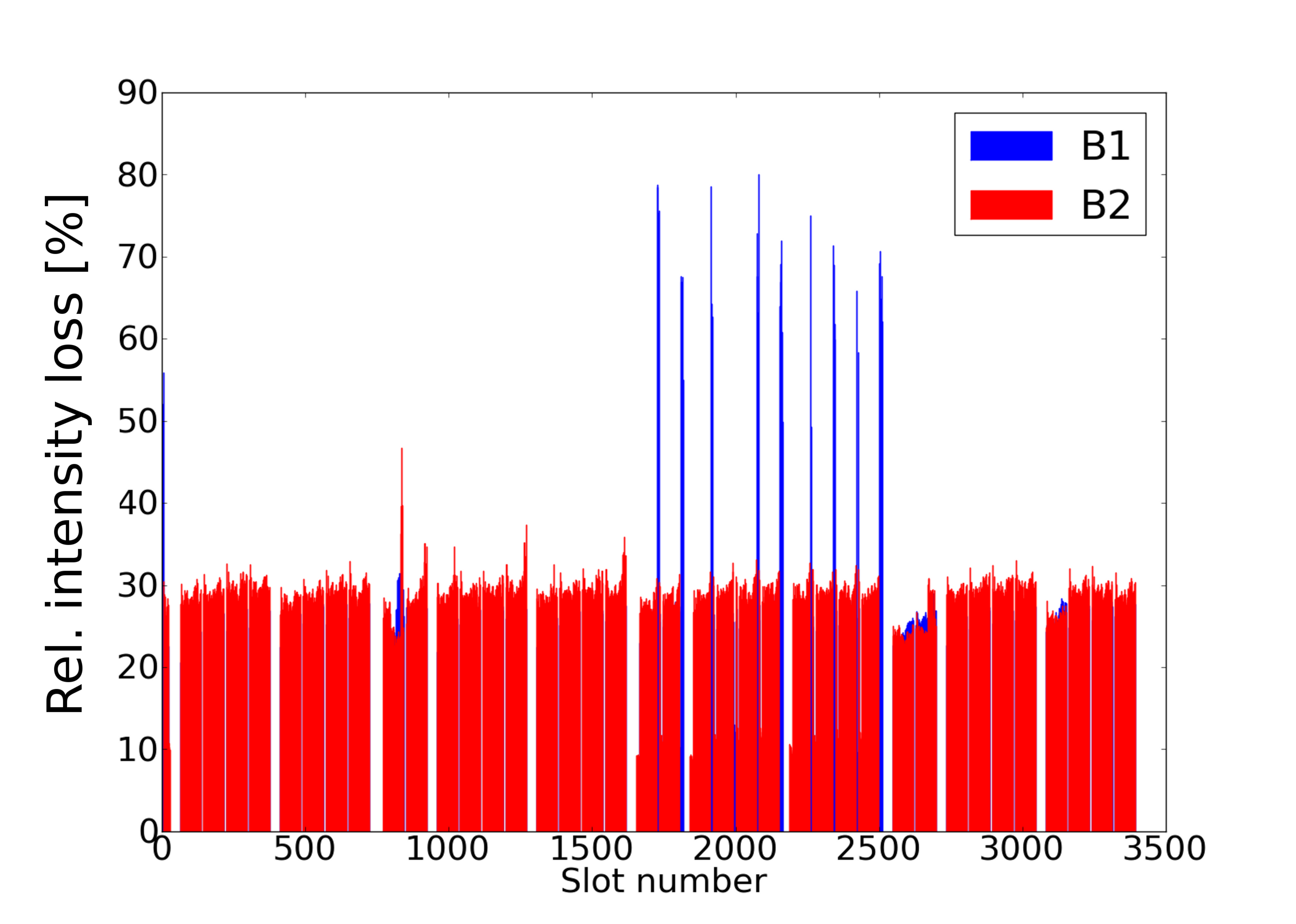}
\label{fig-snowflake selection}
}
\caption{Measured bunch intensities during luminosity production of fill 2646.}
\label{fig-snowflake}
\end{figure}

\begin{figure}
 \centering
\subfloat[Luminosity per IP]{
\includegraphics[width=0.7\linewidth]{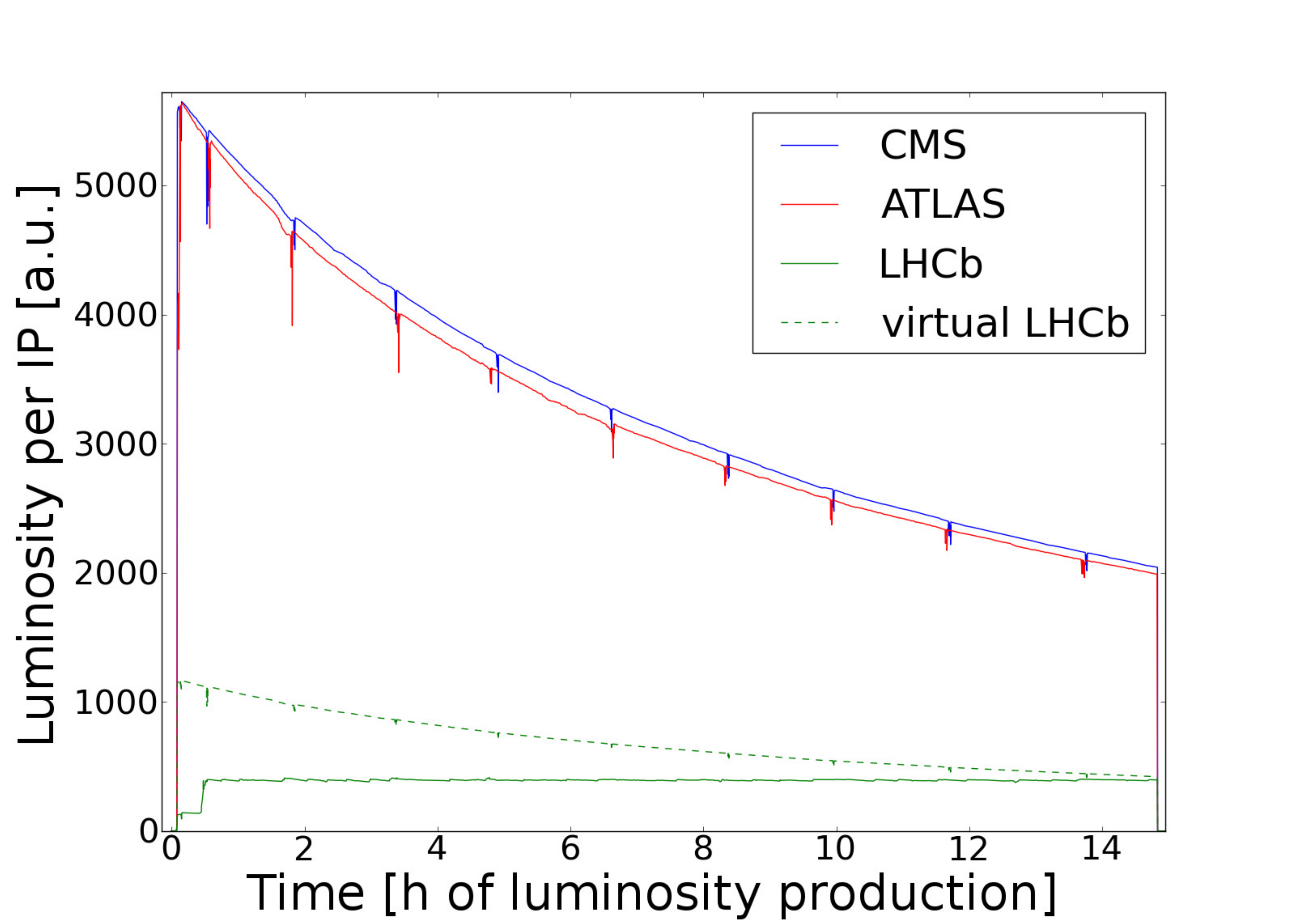}
}
\qquad
\subfloat[Transverse separation in IP8]{
\includegraphics[width=0.7\linewidth]{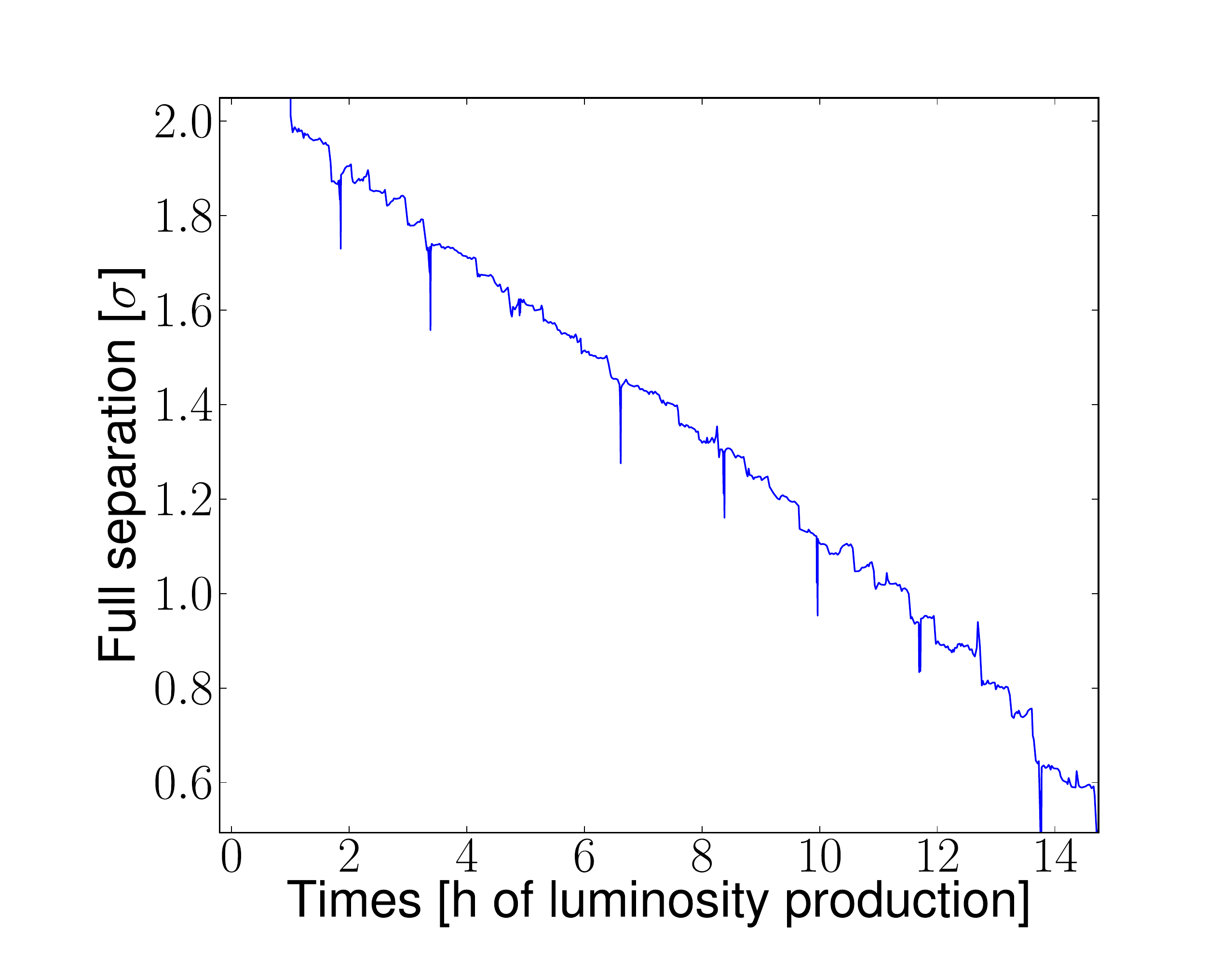}
\label{fig-snowflake sep}
}
\caption{Measured luminosities during luminosity production of fill 2646. The virtual luminosity of IP8 is computed from luminosity in IP1\&5, the resulting reduction factor is used to compute the full separation in IP8.}
\label{fig-snowflake lumi}
\end{figure}

During the 2012 run of the LHC the luminosity was levelled with a transverse offset in IP8. While not harmful for most bunches having HOBB collision in IP1\&5, this technique turned out to be critical for bunches without head-on collision. Indeed, the situation of these bunches is similar to the one described in Fig.~\ref{fig-sdiag}\subref{fig-sdiag sepscan}, however the difference with respect to the process of bringing the beams into collision is that, in this case, the separation is varied in small steps and several minutes are spent at each separation, leaving time for a slow instability to develop. One observation of such an instability is shown in Figs. \ref{fig-snowflake} and \ref{fig-snowflake lumi}. In particular, when comparing the time at which the instabilities occurred (Fig.~\ref{fig-snowflake}\subref{fig-snowgflake BCT}) with the separation computed from measured luminosities (Fig.~\ref{fig-snowflake lumi}\subref{fig-snowflake sep}), it appears that the full separation in IP8 at the time of the instabilities was between 0.9 and 1.6$\sigma$, consistent with the critical separations discussed previously. As can be seen in Fig.~\ref{fig-snowflake}\subref{fig-snowflake selection}, the bunches colliding only in IP8 were located at the end of SPS trains, they were consequently PACMAN bunches, in other words they have a different number of LRBB. Moreover there is bunch-to-bunch variation of the intensity and emittances, which explains why different bunches became unstable at different separations. It is, however, difficult to make quantitative comparison with predications for each individual bunch as many critical parameters are not known to a sufficient precision, in particular the emittances.
\section{Conclusion}
Stability diagrams corresponding to different operational phases of the LHC were derived. It was found that a compromise has to be made when choosing the polarity of the octupoles, the negative polarity providing a better stability at the beginning of the squeeze that degrades during the squeeze due to a partial compensation of the tune spread due to LRBB, as opposed to the positive polarity, which gives less stability at the beginning of the squeeze but rather increases during the squeeze. This effect could not, however, explain instabilities arising at the end of the squeeze, observed in the 2012 run of the LHC with both polarities.

It has been demonstrated that there exists a critical separation, in the order of 1$\sigma$, for which the stability diagram can be dramatically reduced. Observations of coherent instabilities while bringing the beams into collision and during luminosity levelling with a transverse offset are consistent with this effect.

HOBB tune spread is not only larger than the one provided by octupoles or LRBB, it is also dominant on the beam core, rather than the tails, which results in significantly larger stability diagrams. The effect of HOBB could be used to ensure the stability of all bunches in most configurations, in particular by going through the squeeze with colliding beams and ensuring at least one HOBB collision per bunch.

\section{Acknowledgements}
The authors would like to acknowledge G. Arduini, N. Mounet, S. White, B. Salvant, E. M\'etral, S. Redaelli and J. Wenninger for fruitful collaborations and discussions, as well as the LHC-OP crew, in particular G. Papotti and R. Giachino for their work.

\end{document}